%% file: photon_630.tex
\documentstyle[prl,epsfig,eqsecnum,aps,floats]{revtex}
\newcommand{\ppbar}{\mbox{$p\bar{p}$} \ }

\newcommand{\DO}{D\O\ }

\def\ETmiss{\mbox{${\hbox{$E$\kern-0.5em\lower-.1ex\hbox{/}\kern+0.15em}}_{\rm T}$}}

\def\simge
{\mathrel{\rlap{\raise 0.53ex \hbox{$>$}}{\lower 0.53ex \hbox{$\sim$}}}}
\def\simle
{\mathrel{\rlap{\raise 0.4ex \hbox{$<$}}{\lower 0.72ex \hbox{$\sim$}}}}

\begin{document}
\title {Ratio of Isolated Photon Cross Sections at 
{\boldmath $\sqrt{s}=630$} GeV 
and 1800 GeV}
\input{list_of_authors.tex }
\date{16 Aug 2001}
\maketitle

\begin{abstract}
The inclusive cross section for production of isolated photons
has been measured in \ppbar collisions at $\sqrt{s} = 630$ GeV
with the \DO detector at the Fermilab Tevatron Collider. 
The photons span a transverse energy ($E_T$) range 
from 7--49 GeV and have pseudorapidity $|\eta| < 2.5$.
This measurement is combined with the previous \DO result 
at $\sqrt{s} = 1800$ GeV to form a ratio of the cross sections.  
Comparison of next-to-leading-order QCD with the 
measured cross section at 630 GeV and the ratio of cross sections 
show satisfactory agreement in most of the $E_T$ range.  
 
\end{abstract}


\twocolumn

Within the framework of 
Quantum Chromodynamics (QCD), isolated single photons are
direct photons: produced from the primary parton-parton interactions.
Because the dominant production mechanism 
for photons of modest transverse energy ($E_T$) at 
the Fermilab Tevatron is gluon Compton
scattering ($qg \rightarrow \gamma q$), the cross section for 
direct-photon production is sensitive to the gluon distribution in the 
proton \cite{gcomp}.
A measurement of the final state photons provides a probe of 
QCD without additional complications from fragmentation and 
jet identification, providing a powerful and effective means 
for studying the constituents of hadronic matter.
 
Previous experiments, at center-of-mass energies of both 630 GeV
\cite{UA2} and 1800 GeV \cite{{D018},{CDF18}}, 
have reported photon production in excess of next-to-leading-order (NLO) QCD
predictions at low transverse energies ($E^{\gamma}_T \simle 30$ GeV).
This disagreement with data could result from gluon radiation not
included in NLO calculations \cite{huston} 
or because the parton distributions are not well known \cite{pdf}.

In this Letter, we present a measurement of the isolated
photon cross section in  \ppbar collisions
for photons in two pseudorapidity regions, $|\eta| < 0.9$ and 
$1.6 < |\eta | < 2.5$, 
where $\eta = -\ln \tan \frac{\theta}{2}$ and $\theta$  
is the polar angle with respect to the proton beam.  
We compare the production cross section at $\sqrt{s} = 630$ GeV with the 
previously published \DO results at $\sqrt{s} = 1800$ GeV \cite{D018}.
A ratio of the cross sections at different energies 
reduces systematic uncertainties and minimizes
the sensitivity to the choice of parton distribution functions (PDF)
because the measurements at both energies use the same detector
and the same analysis method.

The cross section measurement 
at 630 GeV uses a sample of 520 nb$^{-1}$ of data
recorded in 1995~\cite{lum} with the \DO detector  
at the Fermilab Tevatron~\cite{D0}.  The analysis
uses the uranium/liquid argon calorimeter to identify electromagnetic (EM) 
showers, and the drift chambers in front of the calorimeter 
to differentiate photon showers
from electron showers.  The EM calorimeter provides
full azimuthal ($\phi$) coverage, and consists
of a central cryostat (CC) with
$|\eta | \simle 1.1$, and two forward cryostats (EC) with
$1.4 \simle |\eta | \simle 4.0$. The EM calorimeter is
divided into four longitudinal layers, EM1--EM4,
of approximately 2, 2, 7, and 10 radiation lengths, respectively.
The EM  energy resolution in the central and forward calorimeter is 
given by $\sigma_E/E = \{15\%/\sqrt{E\rm{(GeV)}}\} \oplus 0.3 \%$.

Photons interacting in the calorimeter are detected using a three-level
triggering system.  The first level consists of scintillation counters near
the beam pipe, which detect inelastic \ppbar collisions.  The second
level requires a minimum energy deposition in a $\Delta \phi \times 
\Delta \eta = 0.2 \times 0.2$ trigger tower, with thresholds 
of 2.0, 3.0, and 7.0 GeV.  In the final step, calorimeter clusters 
are formed with corresponding thresholds of 4.5, 8.0, and 14.0 GeV.  
The trigger efficiency is determined for the 14.0 and 8.0 GeV
thresholds by taking the ratio of events passing each 
trigger criteria to those passing the 8.0 and 4.5 GeV criteria, 
respectively, in an energy regime where the lower 
threshold trigger is 100\% efficient.
Monte Carlo studies of the trigger algorithms show
agreement with the data for the two higher energy triggers, and are
used to determine the trigger efficiency for the 4.5 GeV
trigger.  Trigger efficiencies are typically about 20\% at the nominal
energy threshold and rise to almost 100\% a few GeV above the
threshold value.  Consequently, photon candidates are accepted only
for transverse energies of at least 7.35, 10, and 16 GeV
for the three triggers, respectively.

Photon candidates are identified as energy clusters
located well within the pseudorapidity boundaries 
of the central calorimeter or the forward calorimeter,
and, in the central calorimeter, located at 
least 1.6 cm from the azimuthal section boundaries.  The event vertex 
position is required to be within 50 cm of the center of the detector.  
The resulting geometric acceptance is $A=0.622 \pm 0.007$ 
($0.787 \pm 0.007$) in the central (forward) region.  
Candidates must pass a series of selection criteria~\cite{D018}, that
identify the energy cluster as an electromagnetic shower.
The total transverse energy near any candidate cluster
must satisfy an isolation requirement 
$E_{T}^{{\cal R} \leq 0.4} - E_{T}^{{\cal R} \leq 0.2} < 2.0$ GeV,
where ${\cal R} = \sqrt{(\Delta \eta)^2 + (\Delta \phi)^2}$ is the
distance from the cluster center.
The combined selection and isolation efficiency, $\epsilon_s$,
is estimated as a function of $E^{\gamma}_T$
from a {\sc{geant}}-based Monte Carlo simulation of the \DO detector.
We find $\epsilon_s \sim 60\%$ (75\%) in the CC (EC) at 8.0 GeV
and $\epsilon_s \sim 88\%$ (90\%) above 20 GeV.
To minimize background
from electrons, photon candidates are rejected if any tracks in the 
drift chamber extrapolate to within
a road of width $\Delta \phi \times \Delta \theta = 0.2 \times 0.2$
defined by the angle subtended by the candidate
photon cluster and the initial interaction vertex. 
The total charged tracking efficiency
is estimated from $Z \rightarrow e^+e^-$ events
to be $0.858 \pm 0.013$ ($0.593 \pm 0.079$) in 
the central (forward) region. 

The predominant background to direct photon production arises
from the decay of $\pi ^0$ or $\eta$ mesons to two photons.    
The fraction of direct photons is determined from the 
energy ($E_1$) deposited in the innermost
longitudinal section of the calorimeter, EM1.  
Photons have a small probability of showering in the
material in front of the calorimeter and, thus, 
tend to deposit little energy in EM1.  Sensitivity to the
amount of EM1 energy can be used to distinguish multiple photon background
from a single photon signal. 
We use the function $f(E_1)=\log_{10}[1+\log_{10}\{1+E_1{\rm{(GeV)}}\}]$
as our discriminant to determine the single photon purity.
The expected distributions of this function for signal and background
are found from events simulated with the
{\sc{pythia}} Monte Carlo~\cite{pythia} and overlaid 
with data acquired using a random trigger to model noise,
pileup, and multiple \ppbar interactions.
Three categories of fully simulated events are generated: those
containing photons, and background events with and without
charged tracks pointing from the interaction vertex to the
EM cluster.  The two different background
samples are generated so that charged and neutral
background fractions can be separately fit to the data, thus
minimizing uncertainties from the tracking efficiency and from 
the model used for jet fragmentation. 
A systematic uncertainty in modeling jet fragmentation
is estimated by varying the multiplicity of neutral mesons in the
core of {\sc{pythia}} jets by $\pm 10 \%$.
The detector response is modeled using a detailed 
{\sc{geant}} simulation with
the energy response in EM1 calibrated to match the data 
from $W \rightarrow e \nu$ events.

\begin{figure}[htbp]
\vspace{-.2cm}
\centerline{\psfig{figure=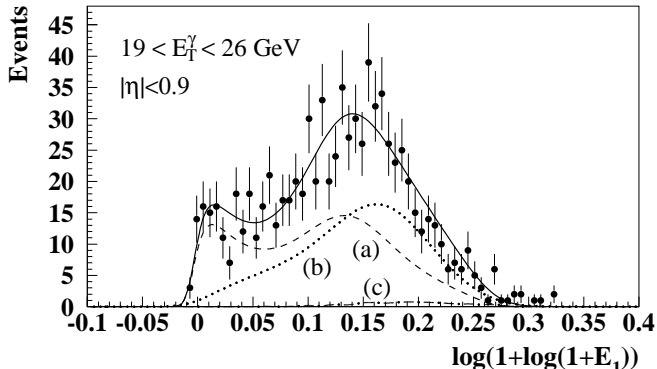,width=3.5in}}
\caption{Distribution of the discriminant, $f$, for determining photon purity, 
where $E_1$ is in units of GeV.  Points with error bars indicate data.  
Broken lines indicate simulated distributions of (a) single photons, 
and jet background (b) without and (c) with charged tracks. 
The solid line depicts a fit sum of all three distributions.} 
\label{fig:fit}
\end{figure}
\begin{figure}[htbp]
\vspace{-1.5cm}
\centerline{\psfig{figure=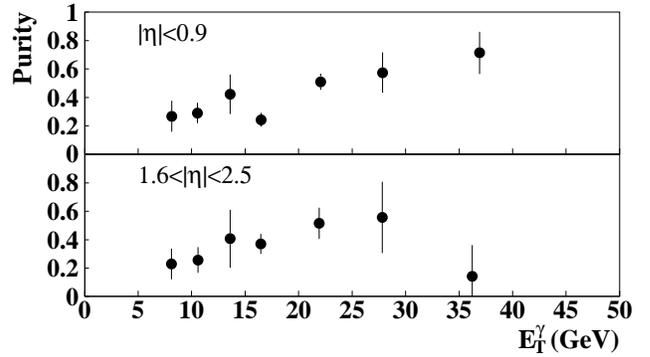,width=3.5in}}
\caption{The photon purity as a function of $E_{T}^{\gamma}$ 
for central and forward photons.  The error bars indicate the
uncertainty in the fit purity and are larger than errors derived
from statistical analysis alone.}
\label{fig:pur}
\end{figure}

The same criteria used to select photon candidates in the
data are applied to the Monte Carlo events.  
The distribution of $f$ from the data is fitted to
a normalized linear combination of Monte Carlo photons and background
with and without charged tracks in the road pointing back to the 
interaction vertex.  The fit is performed in different $E_T^\gamma$
regions using the {\sc{cernlib}} fitting package {\sc{hmcmll}}~\cite{mcmll}, 
with the fractions of
signal and background constrained to be between 0.0 and 1.0.
The purity is defined as the fraction of Monte
Carlo photons in the normalized fitted distribution.  A representative
fit is shown in Fig.~\ref{fig:fit} and the photon purity 
as a function of $E_T^\gamma$ is plotted in Fig.~\ref{fig:pur}.

\begin{figure}[!htbp]
\vspace{-0.8cm}
\centerline{\psfig{figure=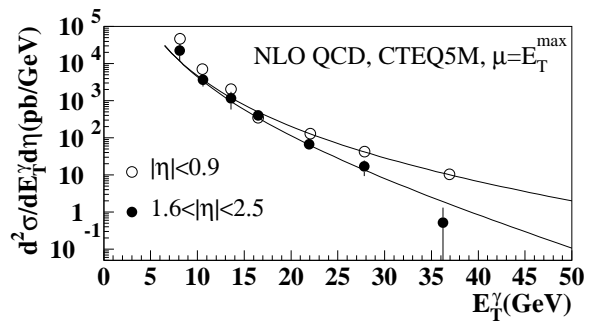,width=3.5in}}
\caption{The cross section for production of isolated photons in
central and forward regions at $\sqrt{s}=630$ GeV.  
The error bars show the total
uncorrelated error, and the curves show cross
sections predicted from NLO QCD.}
\label{fig:xsec}
\end{figure}

\begin{table}[!htbp]
\begin{center}
\begin{tabular}{cccccc}
$E_T^\gamma$ Range & Plotted $ E_T^\gamma $ &
\multicolumn{2}{c}{$ d{^2}\sigma/dE_T^\gamma d\eta$ (pb/GeV)} &
$ \delta \sigma_U $ & $\delta \sigma_C$ \\
(GeV) & (GeV) & Measured & NLO QCD & (\%) & (\%) \\ 
\hline 
\\[-.2cm]
\multicolumn{6}{c}{$|\eta| < 0.9$} \\
~7.35--~9.1 & 8.2  & 47000 & 11400 & 43 & 52 \\ 
~9.1--12.6 & 10.5 & 7160 & 3610 & 26 & 36 \\
12.6--14.7 & 13.6 & 2040 & 1200 & 33  & 25 \\
14.7--18.9 & 16.5 & 351 & 487 & 22 & 19 \\
18.9--26.25 & 22.1 & 131 & 129 & 11  & 13 \\
26.25--29.75 & 27.9 & 42.6 & 41.4 & 25  & 10 \\
29.75--49.0 & 36.9 & 10.5 & 9.95 & 21  & 7 \\ [.15cm]
\multicolumn{6}{c}{$1.6 < |\eta| < 2.5$} \\ 
~7.35--~9.1 & 8.1  & 22400 & 11200 & 48 & 42 \\ 
~9.1--12.6 & 10.6 & 3700 & 3310 & 35 & 31 \\
12.6--14.7 & 13.6 & 1170 & 964 & 50 & 24 \\
14.7--18.9 & 16.5 & 403 & 338 & 20 & 21 \\
18.9--26.25 & 21.9 & 67.3 & 65.4 & 22 & 17 \\
26.25--29.75 & 27.8 & 16.9 & 13.6 & 45 & 16 \\
29.75--49.0 & 36.2 & 0.522 & 1.91 & 160 & 15 \\
\end{tabular}
\end{center}
\caption{The measured and predicted isolated photon 
production cross section at $\sqrt{s}=630$ GeV.
The value for the column labeled ``Plotted $E_T^\gamma $'' is 
determined according to Ref.~\protect\cite{plotet}. 
The columns labeled $\delta \sigma_U$ and $\delta \sigma_C$ 
are, respectively, the uncorrelated and correlated uncertainties.
}
\label{table:xsec}
\end{table}

The final cross sections
$d^2 \sigma / d E^{\gamma}_T d \eta$, after applying efficiency 
and purity corrections, are shown
in Fig.~\ref{fig:xsec} and tabulated in Table~\ref{table:xsec}.  
The error bars show all uncorrelated uncertainties,
which include the statistical uncertainty, and uncertainties from 
selection criteria, trigger efficiency, and
the fitted photon purity.  The 
contribution from the fit to photon 
purity is the largest source of uncorrelated uncertainty. The
correlated uncertainty consists of the uncertainties in luminosity,
tracking efficiency, geometric acceptance, calorimeter
energy scale, and the largest contribution, that from the fragmentation model.

\begin{figure}[!htbp]
\vspace{-1.4cm}
\centerline{\psfig{figure=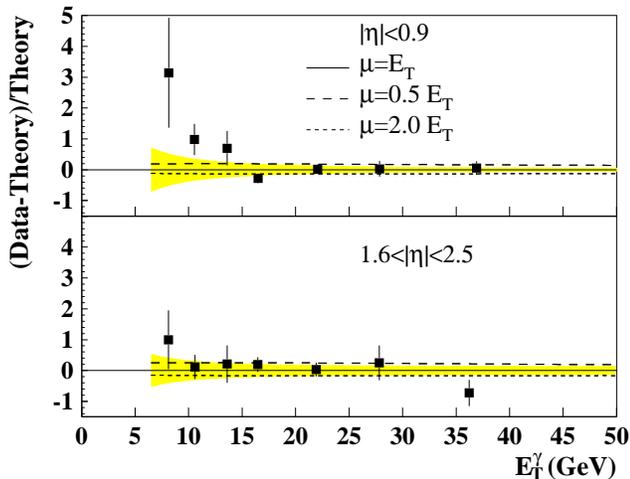,width=3.5in}}
\caption{Comparison of the measured cross section for production
of isolated photons at $\sqrt{s}=630$ GeV
with NLO QCD using CTEQ5M parton distribution functions.
The error bars indicate the
uncorrelated uncertainty and the shaded bands indicate the 
correlated uncertainty }
\label{fig:norm}
\end{figure}

The results are compared with NLO QCD calculations using CTEQ5M
parton distributions~\cite{owens}, with renormalization and factorization 
scales $\mu_R = \mu_F = E_T^{\rm{max}}$, where  
$E_T^{\rm{max}}$ is the maximum photon transverse energy in the event.
Figure~\ref{fig:norm} compares the data and theory.
A covariance matrix $\chi^2$  provides a measure
of the probability that the theory describes the data.  A complete
covariance matrix, composed of correlated and uncorrelated uncertainties,
is determined and the theoretical cross section is compared to
the data with a $\chi^2$ value of 11 (4.6) for 7 degrees of
freedom in the CC (EC) region.  This gives a standard $\chi^2$ probability 
that the theory is consistent with the data at 12\% (71\%) 
probability in the CC (EC) regions.  Deviations between theory
and data are largest at low $E^{\gamma}_T$ in the central region.
These results are in qualitative agreement with those previously
published at $\sqrt{s}=1800$ GeV, where the theory is lower than
the data at low $E^{\gamma}_T$ ($\approx$10--40 GeV) in the CC, 
but is consistent with the data over 
all $E^{\gamma}_T$ in the EC \cite{D018}.  Using different PDFs
changed the cross section by less than 5\%~\cite{pdfs}.  Setting scales to
$\mu_R = \mu_F = 2.0 E_T^{\rm{max}}$ or $\mu_R = \mu_F = 0.5 E_T^{\rm{max}}$
changed the cross section by about 20\%, as shown in Fig.~\ref{fig:norm}.

\begin{figure}[htbp]
\vspace{-1.5cm}
\centerline{\psfig{figure=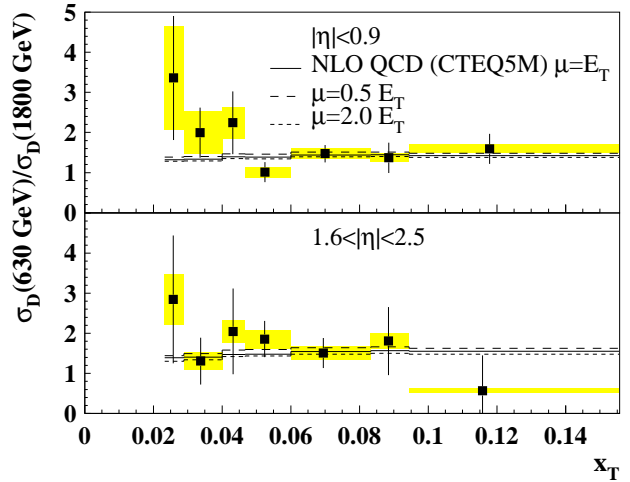,width=3.5in}}
\caption{The ratio of the dimensionless cross 
sections, 
$\sigma_D(\sqrt{s}=630{\rm{~GeV}})/\sigma_D(\sqrt{s}=1800{\rm{~GeV}})$.
The error bars indicate the
uncorrelated uncertainty and the shaded bands indicate the 
correlated uncertainty.}
\label{fig:ratio}
\end{figure}

In the simple parton model, the dimensionless cross section
$E_T^4 \cdot E \frac{d^3\sigma}{dp^3}$, as a function of
$x_T = \frac{2 E_T}{\sqrt{s}}$, is independent of $\sqrt{s}$.   
Although deviations from such naive scaling are expected,
the dimensionless framework provides a useful context for 
comparison with QCD.  The experimental
dimensionless cross section, averaged over azimuth, becomes
$\sigma_D=\frac{E_T^3}{2\pi} \cdot d^2\sigma/dE_T d\eta$.
The ratio 
$\sigma_D(\sqrt{s}=630 {\rm{~GeV}})/\sigma_D(\sqrt{s}=1800 {\rm{~GeV}})$
is determined by combining the cross section
reported in this Letter with the \DO measurement 
at $\sqrt{s} = 1800$ GeV~\cite{{D018},{E811}}.
The ratio is shown as a function of $x_T$
in Fig.~\ref{fig:ratio} and Table~\ref{table:ratio} together with
the NLO QCD prediction.



\begin{table}[htbp]
\begin{center}
\begin{tabular}{cccccc}
$x_T$ Range & Plotted $x_T$ & Ratio & Theory & 
$ \delta \sigma_U (\%) $ & $\delta \sigma_C$ (\%) \\ 
\hline 
\\[-.2cm]
\multicolumn{6}{c}{$|\eta| < 0.9$} \\
0.023--0.029 & 0.026  & 3.36 & 1.32 & 46 & 39 \\ 
0.029--0.040 & 0.034 & 2.00 & 1.34 & 31 & 27 \\
0.040--0.047 & 0.043 & 2.24 & 1.40 & 35  & 18 \\
0.047--0.060 & 0.053 & 1.01 & 1.39 & 25 & 13 \\
0.060--0.083 & 0.070 & 1.47 & 1.44 & 15  & 9 \\
0.083--0.094 & 0.089 & 1.37 & 1.45 & 27  & 8 \\
0.094--0.156 & 0.118 & 1.59 & 1.42 & 23  & 7 \\ [.15cm]
\multicolumn{6}{c}{$1.6 < |\eta| < 2.5$} \\ 
0.023--0.029 & 0.026  & 2.84 & 1.39 & 56 & 22 \\ 
0.029--0.040 & 0.034 & 1.31 & 1.41 & 45 & 17 \\
0.040--0.047 & 0.043 & 2.05 & 1.47 & 52  & 14 \\
0.047--0.060 & 0.052 & 1.86 & 1.48 & 24 & 13 \\
0.060--0.083 & 0.070 & 1.51 & 1.54 & 24  & 11 \\
0.083--0.094 & 0.088 & 1.81 & 1.57 & 47  & 11 \\
0.094--0.156 & 0.116 & 0.563 & 1.55 & 160 & 10 \\
\end{tabular}
\end{center}
\caption{The measured ratio and NLO QCD prediction for the dimensionless
cross section at $\sqrt{s}=630$ GeV to that at $\sqrt{s}=1800$ GeV.  
The columns labeled $\delta \sigma_U$ and $\delta \sigma_C$ are 
the uncorrelated and correlated uncertainties, respectively.
}
\label{table:ratio}
\end{table}

Comparison of the theoretical cross section ratio
to the data, using the complete covariance matrix,  
gives a $\chi^2$ value of 6.5 (3.0) for 7 degrees of
freedom in the CC (EC), which corresponds to a standard $\chi^2$ 
probability of 49\% (89\%) in the CC (EC) region.
Although the lowest $x_T$ points are systematically higher
than NLO QCD predictions in both the CC and EC regions, 
the deviations are not significant in light of our combined
statistical and systematic uncertainties, and there 
exists good agreement between the measured ratio and theory.

We have measured the production cross section for
isolated photons in $p \overline{p}$ collisions at $\sqrt{s}=630$ GeV
and compared this cross section with that measured at $\sqrt{s}=1800$ GeV.
The measurement is higher than the 
theoretical prediction at low $E_T$ in the
central rapidity region but agrees at all other $E_T$ and in
the forward rapidity region.   The difference between data
and theory is less significant 
for the ratio of cross sections, and the theory is 
consistent with the data over all $E_T$.

\input{acknowledgement_paragraph}
\end{document}

%% file: list_of_authors.tex
%
\author{                                                                      
V.M.~Abazov,$^{23}$                                                           
B.~Abbott,$^{57}$                                                             
A.~Abdesselam,$^{11}$                                                         
M.~Abolins,$^{50}$                                                            
V.~Abramov,$^{26}$                                                            
B.S.~Acharya,$^{17}$                                                          
D.L.~Adams,$^{59}$                                                            
M.~Adams,$^{37}$                                                              
S.N.~Ahmed,$^{21}$                                                            
G.D.~Alexeev,$^{23}$
% ...
\\                                                                            
\vskip 0.30cm                                                                 
\centerline{(D\O\ Collaboration)}                                             
\vskip 0.30cm                                                                 
}                                                                             
\address{                                                                     
\centerline{$^{1}$Universidad de Buenos Aires, Buenos Aires, Argentina}       
\centerline{$^{2}$LAFEX, Centro Brasileiro de Pesquisas F{\'\i}sicas,         
                  Rio de Janeiro, Brazil}                                     
\centerline{$^{3}$Universidade do Estado do Rio de Janeiro,                   
                  Rio de Janeiro, Brazil}                                     
\centerline{$^{4}$Institute of High Energy Physics, Beijing,                  
                  People's Republic of China}                                 
\centerline{$^{5}$Universidad de los Andes, Bogot\'{a}, Colombia}             
\centerline{$^{6}$Charles University, Center for Particle Physics,            
                  Prague, Czech Republic}                                     
\centerline{$^{7}$Institute of Physics, Academy of Sciences, Center           
                  for Particle Physics, Prague, Czech Republic}               
\centerline{$^{8}$Universidad San Francisco de Quito, Quito, Ecuador}         
\centerline{$^{9}$Institut des Sciences Nucl\'eaires, IN2P3-CNRS,             
                  Universite de Grenoble 1, Grenoble, France}                 
\centerline{$^{10}$CPPM, IN2P3-CNRS, Universit\'e de la M\'editerran\'ee,     
                  Marseille, France}                                          
\centerline{$^{11}$Laboratoire de l'Acc\'el\'erateur Lin\'eaire,              
                  IN2P3-CNRS, Orsay, France}                                  
\centerline{$^{12}$LPNHE, Universit\'es Paris VI and VII, IN2P3-CNRS,         
                  Paris, France}                                              
\centerline{$^{13}$DAPNIA/Service de Physique des Particules, CEA, Saclay,    
                  France}                                                     
\centerline{$^{14}$Universit{\"a}t Mainz, Institut f{\"u}r Physik,            
                  Mainz, Germany}                                             
\centerline{$^{15}$Panjab University, Chandigarh, India}                      
\centerline{$^{16}$Delhi University, Delhi, India}                            
\centerline{$^{17}$Tata Institute of Fundamental Research, Mumbai, India}     
\centerline{$^{18}$Seoul National University, Seoul, Korea}                   
\centerline{$^{19}$CINVESTAV, Mexico City, Mexico}                            
\centerline{$^{20}$FOM-Institute NIKHEF and University of                     
                  Amsterdam/NIKHEF, Amsterdam, The Netherlands}               
\centerline{$^{21}$University of Nijmegen/NIKHEF, Nijmegen, The               
                  Netherlands}                                                
\centerline{$^{22}$Institute of Nuclear Physics, Krak\'ow, Poland}            
\centerline{$^{23}$Joint Institute for Nuclear Research, Dubna, Russia}       
\centerline{$^{24}$Institute for Theoretical and Experimental Physics,        
                   Moscow, Russia}                                            
\centerline{$^{25}$Moscow State University, Moscow, Russia}                   
\centerline{$^{26}$Institute for High Energy Physics, Protvino, Russia}       
\centerline{$^{27}$Lancaster University, Lancaster, United Kingdom}           
\centerline{$^{28}$Imperial College, London, United Kingdom}                  
\centerline{$^{29}$University of Arizona, Tucson, Arizona 85721}              
\centerline{$^{30}$Lawrence Berkeley National Laboratory and University of    
                  California, Berkeley, California 94720}                     
\centerline{$^{31}$University of California, Davis, California 95616}         
\centerline{$^{32}$California State University, Fresno, California 93740}     
\centerline{$^{33}$University of California, Irvine, California 92697}        
\centerline{$^{34}$University of California, Riverside, California 92521}     
\centerline{$^{35}$Florida State University, Tallahassee, Florida 32306}      
\centerline{$^{36}$Fermi National Accelerator Laboratory, Batavia,            
                   Illinois 60510}                                            
\centerline{$^{37}$University of Illinois at Chicago, Chicago,                
                   Illinois 60607}                                            
\centerline{$^{38}$Northern Illinois University, DeKalb, Illinois 60115}      
\centerline{$^{39}$Northwestern University, Evanston, Illinois 60208}         
\centerline{$^{40}$Indiana University, Bloomington, Indiana 47405}            
\centerline{$^{41}$University of Notre Dame, Notre Dame, Indiana 46556}       
\centerline{$^{42}$Iowa State University, Ames, Iowa 50011}                   
\centerline{$^{43}$University of Kansas, Lawrence, Kansas 66045}              
\centerline{$^{44}$Kansas State University, Manhattan, Kansas 66506}          
\centerline{$^{45}$Louisiana Tech University, Ruston, Louisiana 71272}        
\centerline{$^{46}$University of Maryland, College Park, Maryland 20742}      
\centerline{$^{47}$Boston University, Boston, Massachusetts 02215}            
\centerline{$^{48}$Northeastern University, Boston, Massachusetts 02115}      
\centerline{$^{49}$University of Michigan, Ann Arbor, Michigan 48109}         
\centerline{$^{50}$Michigan State University, East Lansing, Michigan 48824}   
\centerline{$^{51}$University of Nebraska, Lincoln, Nebraska 68588}           
\centerline{$^{52}$Columbia University, New York, New York 10027}             
\centerline{$^{53}$University of Rochester, Rochester, New York 14627}        
\centerline{$^{54}$State University of New York, Stony Brook,                 
                   New York 11794}                                            
\centerline{$^{55}$Brookhaven National Laboratory, Upton, New York 11973}     
\centerline{$^{56}$Langston University, Langston, Oklahoma 73050}             
\centerline{$^{57}$University of Oklahoma, Norman, Oklahoma 73019}            
\centerline{$^{58}$Brown University, Providence, Rhode Island 02912}          
\centerline{$^{59}$University of Texas, Arlington, Texas 76019}               
\centerline{$^{60}$Texas A\&M University, College Station, Texas 77843}       
\centerline{$^{61}$Rice University, Houston, Texas 77005}                     
\centerline{$^{62}$University of Virginia, Charlottesville, Virginia 22901}   
\centerline{$^{63}$University of Washington, Seattle, Washington 98195}       
}                                                                             

%% file: acknowledgement_paragraph.tex
%
We thank W.~Vogelsang and J.F.~Owens for their assistance
with the theoretical calculations.
We thank the staffs at Fermilab and collaborating institutions, 
and acknowledge support from the 
Department of Energy and National Science Foundation (USA),  
Commissariat  \` a L'Energie Atomique and 
CNRS/Institut National de Physique Nucl\'eaire et 
de Physique des Particules (France), 
Ministry for Science and Technology and Ministry for Atomic 
   Energy (Russia),
CAPES and CNPq (Brazil),
Departments of Atomic Energy and Science and Education (India),
Colciencias (Colombia),
CONACyT (Mexico),
Ministry of Education and KOSEF (Korea),
CONICET and UBACyT (Argentina),
The Foundation for Fundamental Research on Matter (The Netherlands),
PPARC (United Kingdom),
Ministry of Education (Czech Republic),
and the A.P.~Sloan Foundation.

%% file: photon_630.bbl
\begin{thebibliography}{99}
\vspace{-1.6 cm} 
%
\bibitem[*]{lehner}
Visitor from University of Zurich, Zurich, Switzerland.
%
\vskip 0.25cm

\bibitem{gcomp} J.~Owens, Rev. Mod. Phys. {\bf 59}, 465 (1987). 
\bibitem{UA2} J. Alitti {\it et al.} (UA2 Collaboration),  Phys.~Lett.~B 
{\bf 263}, 544 (1991);  R. Ansari {\it et al.} (UA2 Collaboration),  
Z. Phys. C {\bf 41}, 395 (1988).
\bibitem{D018}  B.~Abbott {\it et al.} (\DO Collaboration) Phys.~Rev.~Lett. 
{\bf 84}, 2786 (2000).
\bibitem{CDF18} F. Abe {\it et al.} (CDF Collaboration),  Phys.~Rev.~Lett. 
{\bf 73}, 2662 (1994); F. Abe {\it et al.} (CDF Collaboration), Phys.~Rev.~D 
{\bf 48}, 2998 (1993). 
\bibitem{huston} J.~Huston {\it et al.}, Phys.~Rev.~D {\bf 51}, 6139 (1995);
H.-L.~Lai and H.-N. Li, {\sl ibid.} {\bf 58}, 114020 (1998);
L.~Apanasevich {\it et al.}, {\sl ibid.} {\bf 59}, 074007 (1999).
\bibitem{pdf} M. Gl\"{u}ck, {\it et al.}, Phys.~Rev.~Lett. {\bf 73}, 
388 (1994);
W Giele, in {\sl Proceedings of the 5th International 
Symposium on Radiative Corrections}, 
Carmel, CA, September 2000 (to be published electronically at   
http://www.slac.stanford.edu/econf/).
\bibitem{lum} J.~Krane, J.~Bantley, and D.~Owen, Fermilab-TM-2000, (1997) unpublished. 
\bibitem{D0} S. Abachi {\it et al.} (\DO Collaboration),  
Nucl.~Instrum.~Methods~Phys.~Res.~A {\bf 338}, 185 (1994).
\bibitem{pythia} T. Sj\"{o}strand, Comput.~Phys.~Commun.~{\bf 82}, 74 (1995).
\bibitem{mcmll} R.~Barlow and C.~Beeston, Comput.~Phys.~Commun.~{\bf 77}, 
219 (1993).
\bibitem{owens} H.~Baer, J.~Ohnemus, and J.F.~Owens, Phys.~Rev.~D {\bf 42}, 
61 (1990); W. Vogelsang and A. Vogt, Nucl.~Phys.~B {\bf 453}, 334 (1995).
The authors of these predictions have verified that their calculated
cross sections are consistent over the range of photon transverse energies
measured here.
\bibitem{plotet} G.D.~Lafferty and T.R.~Wyatt, Nucl.~Instrum.~Methods
Phys.~Res.~A {\bf 355}, 541 (1995).
\bibitem{pdfs} CTEQ5M, CTEQ5HJ, MRST, MRSTg$\uparrow$, and 
MRSTg$\downarrow$ were compared.  For MRST, see
A.D.~Martin {\it et al.} Eur.~Phys.~J.~C {\bf 14}, 133 (2000). 
\bibitem{E811} The cross section at 1800 GeV is increased by 3.4\%
from the published result to account for the current value 
of the proton-antiproton total cross section as per
B.~Abbott {\it et al.}
(\DO Collaboration), Phys.~Rev.~D {\bf 61}, 072001 (2000);
C. Avila {\it et al.} (E811 Collaboration)
Phys.~Lett.~B {\bf 445}, 419 (1999).

\end{thebibliography}
